\documentclass[12pt]{article}

\usepackage[dvips]{epsfig,graphicx, color}

\usepackage{times}

\newenvironment{sciabstract}{%
\begin{quote} \bf}
{\end{quote}}

\newcounter{lastnote}

\title{Integrin activation - the importance of a positive feedback}

\author
{Dagmar Iber\footnote{To whom correspondance should be addressed:
Mathematical Institute, Centre for Mathematical Biology, 24-29 St.
Giles, Oxford OX1 3LB, UK
email: iber@maths.ox.ac.uk}$^1$ and Iain D. Campbell$^2$ \\[3mm]
{\normalsize $^1$Mathematical Institute, Centre for Mathematical Biology, 24-29 St. Giles, Oxford OX1 3LB, UK;}\\
{\normalsize $^2$ Department of Biochemistry, University of
Oxford, Oxford, OX1 3QU, UK}}

\date{}

\begin{document}


\baselineskip12pt


\maketitle
\newpage
\begin{sciabstract}
Integrins mediate cell adhesion and are essential receptors for
the development and functioning of multicellular organisms.
Integrin activation is known to require both ligand and talin
binding and to correlate with cluster formation but the activation
mechanism and precise roles of these processes are not yet
resolved. Here mathematical modeling, with known experimental
parameters, is used to show that the binding of a stabilizing
factor, such as talin, is alone insufficient to enable
ligand-dependent integrin activation for all observed conditions;
an additional positive feedback
is required.\\
\end{sciabstract}


\newpage

\section*{Introduction}
Integrins, large membrane-spanning heterodimeric proteins, were so
named for their ability to link the extracellular and
intracellular skeletons \cite{Tamkun86}. As an important class of
cell adhesion receptors they participate in a wide-range of
biological interactions, including development, tissue repair,
angiogenesis, inflammation and hemostasis \cite{Horwitz03}. Cell
adhesion and detachment as well as controlled actin polymerisation
inside the cell are of particular importance in cell migration.
The speed of cell movement depends on the density of integrins and
ligands as well as their affinity of binding \cite{Palecek97}.
Integrins are key components of focal adhesions, dynamic
multi-protein complexes that are involved in the regulation of
cell adhesion and migration. Focal adhesions \cite{Zamir01}
provide a physical link between integrins and the actin
cytoskeleton as well as sites for signal transduction into the
cell interior. Information about identified interactions and
players in these complexes is ever-increasing but our overall
understanding of how the ensemble works remains relatively poor.

According to the current model of integrin activation, ligand
binding shifts the equilibrium between different integrin
conformations to the active one \cite{Hynes02}. The two extreme
conformations of this allosteric protein are a bent or 'closed'
conformation which represents the low affinity state for ligand
and an 'open' conformation that will bind with high affinity to
ligand. Conformational changes in the extracellular domain affect
the cytoplasmic tails, which are separated in the open
conformation but not in the closed. Separation of the cytoplasmic
domains promotes their interaction with cytoskeletal and signal
transduction molecules, and thus the activation of integrins and
downstream signaling. The conformational equilibrium can be
influenced both by ligand binding to the extracellular domain
(outside-in signaling) and by binding of cytoplasmic proteins to
the separated cytoplasmic domains (inside-out signaling). As well
as changes in affinity induced by structural changes, integrins
can also modulate their avidity by clustering, thus changing the
valency of their interactions with ligand \cite{Carman03}.

Previous theoretical studies have addressed the mechanism of
integrin clustering \cite{Ward94,Irvine02_Int} but they have not
yet explored whether ligand binding is sufficient or if other
factors such as clustering are necessary for integrin activation.
In principle, ligand-dependent integrin activation can be enabled
by any of the four different mechanisms depicted in Figure
\ref{Fig_model}. A) Extracellular ligand alone is sufficient to
activate integrins and other factors are only important for
downstream signaling (Fig. \ref{Fig_model}A); B) Ligand-dependent
integrin activation requires the binding of an intracellular
stabilizing factor that interacts with and stabilizes the on state
(Fig. \ref{Fig_model}B). C) A positive feedback, provided either
by integrin-ligand pairs themselves or by self-interactions, is
important \cite{Li03}) (Fig. \ref{Fig_model}C); D) A positive
feedback that increases the activity of an intracellular
stabilizing factor is required (Fig. \ref{Fig_model}D). The latter
would need to be based on a larger intracellular signaling
network.

Mathematical modeling is used in the following to evaluate the
physiological potential of these mechanisms. The analysis suggests
that, at least for some physiological conditions, binding of
ligand (even if it is preclustered) and/or a stabilizing factor
(Fig. \ref{Fig_model}A,B) are insufficient for ligand-dependent
integrin activation; a positive feedback (Fig. \ref{Fig_model}C,D)
is required which results in the further stabilization of the
active conformation. Integrin clustering is a likely consequence
of such a positive feedback.

\section*{A model for integrin activation }
Overly simplified, integrins can be taken to exist in one of two
states, a closed low affinity or an open high affinity
conformation. In the absence of ligand the equilibrium is biased
towards the inactive, low affinity conformation \cite{Tadokoro03}.
Binding of ligand ($L$) stabilizes the active, high affinity
conformation. Integrin activation can thus be measured as the
fraction of integrins bound by ligand. The analysis can be greatly
simplified if the fraction of integrins bound to ligand is
determined indirectly from the fraction of ligand bound by
integrins. In this case, integrins in the closed, low affinity
conformation do not have to be considered explicitly. We will
assume that there is a constant number of unbound integrins in the
high affinity conformation since any integrin that binds to ligand
can be replaced rapidly from the integrins in the low affinity
conformation by conformational changes. While ligand binding will
eventually deplete the pool of unbound integrins, this does not
need to be included in an analysis of mechanisms for integrin
signal initiation: if integrin-ligand binding occurs to a level
that integrin depletion becomes relevant the activation mechanism
can be considered successful independently of the exact final
number of integrin-ligand complexes.

Under physiological conditions, the ligand is expected to be
sufficiently dense that spatial details and diffusion constraints
can be neglected in the initial steps of focal adhesion assembly.
Transport of integrins from parts of the cell that are not in
contact with the substrate is not relevant for the initial
activation of cell adhesion signaling cascades and is not
considered in this model.

The model follows the binding of single integrins to ligands; the
probability of integrin-ligand complex formation can be calculated
by solving the appropriate Master equation. The Master equation is
a gain-loss equation for the probabilities of the separate states
with $n$ integrin-ligand complexes \cite{vanKampen}, from which
the macroscopic kinetic equation can be derived when the density
fluctuations are negligible. In order to decide which of the four
mechanisms depicted in Figure \ref{Fig_model} is sufficient for
ligand-dependent integrin activation, the fraction of ligand bound
by integrin is determined for each case.

\subsection*{Ligand engagement is not sufficient for integrin activation}
According to the first mechanism (Fig. \ref{Fig_model}A),
ligand-dependent integrin activation does not require any further
supporting processes and interactions. The Master equation for the
formation of $i$ integrin-ligand complexes is thus given as

\begin{equation}\label{Master_eq_integrin}
\dot{p}_{i} = k_{-} (i+1) p_{i+1} +k_+(L-i+1) p_{i-1}
-(k_{-}i+k_+(L-i))p_{i}
\end{equation}

with $p_{i} = 0$ if $i \notin [0,L]$, where $p_i$ is the
probability that $i$ (open) integrins are bound to a ligand with
$L$ sites. $k_{+},k_{-}$ refer to the on- and off-rate of the
integrins. The on-rates include the density of integrins which are
taken to be constant (see previous section). The first two terms
in (\ref{Master_eq_integrin}) represent the gains of state $i$ due
to the transition from other states, that is due to unbinding of
an integrin from a complex with $i+1$ integrins (first term) or
the binding to a complex with $i-1$ integrins (second term). The
bracketed term is the loss due to transitions from $i$ into other
states either by unbinding or binding of an integrin.

For the steady state ($\dot{p}_{i}=0$) we obtain from
(\ref{Master_eq_integrin})

\begin{equation}\label{steadystate}
p_{i} = \frac{\left(\begin{array}{c} L \\ i \end{array}\right)
\left(\frac{k_+}{k_-}\right)^{i}}{(1+\frac{k_+}{k_-})^L}
\end{equation}

\begin{figure} [!h]
\linespread{1.3}
\begin{center}
\includegraphics[width=7cm]{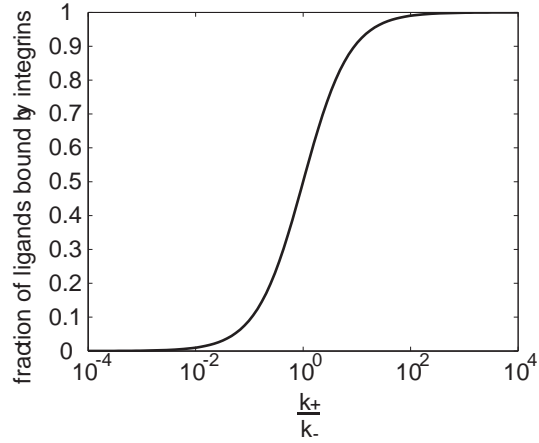}
\caption{\label{Fig_L} {\small \textbf{Ligand engagement is not
sufficient for integrin activation} The integrin saturation of the
ligand, $\frac{<I_b>}{L}$, is plotted against the integrin-ligand
association constant $K_a=\frac{k_+}{k_-}$; L=10. As discussed in
the text, experimental data suggest that $\frac{k_+}{k_-} \ll 1$
such that the fraction of activated integrins would be very small.}}
\end{center}
\end{figure}

The expectation value for the number of integrins bound to a given
ligand with $L$ binding sites is given by  $<I_b> = \sum_{i=0}^{L}
i p_{i} $. Integrin activation therefore requires $p_i \gg p_0$
($i>0$) and thus $\frac{k_+}{k_-} \gg 1$ (Fig. \ref{Fig_L}), which
corresponds to a high on- and low off-rate. This is because the
probability of $i$ bound integrins is proportional to the $i$th
power of $\frac{k_+}{k_-}$ and only for $\frac{k_+}{k_-}
> 1$ this probability increases with increasing $i$, that is with
increasing numbers of integrins bound. \\

$k_- \sim 1$ s$^{-1}$ has been established experimentally
\cite{Vitte04}; the 2-dimensional on-rate is more difficult to
determine and has to be calculated from the 2-dimensional
dissociation constant which itself is difficult to measure and
often obtained via its 3-dimensional counterpart ($K_D^{3d} \sim 6
\times 10^{-8}$ M for the high affinity conformation
\cite{Faull93,Suehiro97} and $K_D^{3d} > 1 \mu$M for the low
affinity conformation \cite{Faull93}). It is generally assumed
that the two dissociation constants are linearly related such that
$K_D^{2d} = \eta K_D^{3d}$. The value of the conversion factor
$\eta$ is, however, still a matter of debate \cite{Moy99}, since
theoretical estimates are not in complete agreement with latest
experimental results. According to theoretical estimates for a
protein complex that spans about 20 nm $\eta = 1.2 \times 10^{19}
\frac{\sharp}{m^2 M}$ \cite{Bell78}. Studies employing
protein-coated beads report $\eta = 8 \times 10^{24}
\frac{\sharp}{m^2 M}$ \cite{Moy99} and $\eta = 10^{22}
\frac{\sharp}{m^2 M}$ \cite{Kuo93}. The lower $\eta$ value is
corroborated by measurements of the 2-dimensional dissociation
constant for the LFA-3/CD2 adhesion pair that were carried out
using cells instead of beads \cite{Dustin96}. While there are
experimental difficulties involved in getting an exact measurement
of this parameter, the discrepancy between the theoretical and
experimental estimates may well reflect a lower affinity of
integrins when membrane bound, possibly due to steric constraints
which will have a profound impact on this allosteric protein. We
will therefore follow Moy and co-workers (1999) and use $\eta =
10^{22} \frac{\sharp}{m^2 M}$ for converting integrin dissociation
constants. For $\eta = 10^{22} \frac{\sharp}{m^2 M}$ we find
$k_{on} \sim 2 \times 10^{-3} \mu$m$^2$ ($\sharp$ s)$^{-1}$. In
order to determine the frequency of integrin-ligand bond formation
the density of open integrins needs to be taken into
consideration. The average total (closed and open) integrin
density on the cell surface has been estimated as $\rho \sim
1-3\times 10^2 \mu m^{-2}$ \cite{Wiseman04}. At least in some
cells (e.g. platelets) more than $95\%$ of all integrins are in
the closed inactive conformation in the absence of ligand
\cite{Tadokoro03}; the density of integrin in the open
conformation is therefore small ($\rho_o < 10 \mu m^{-2}$). Given
that $k_+=k_{on}\rho_o$ this implies, using the experimental
estimate for $\eta$, that $k_+ \sim 10^{-1}-10^{-2}$ s$^{-1}$ such
that $\frac{k_+}{k_-} \ll 1$, which is insufficient to drive
integrin activation. A larger $k_+$ can be achieved if a larger
fraction of integrins is in the open conformation already in the
absence of ligand. However, this leads to ligand-independent
integrin activation. We can conclude that for experimentally
determined parameters the model predicts that
ligand-dependent integrin activation will not occur without further supporting  interactions.\\

\subsection*{A stabilizing factor is not sufficient for ligand-dependent integrin activation}
Proteins that bind and stabilize the active integrin conformation
have been suggested to be important for integrin activation (Fig.
\ref{Fig_model}B). The Master equation derived in the previous
section can be extended to include such factors that stabilize
integrin-ligand binding. While talin is an excellent candidate for
such a protein (and we will therefore call this factor talin in
the following), the analysis is kept general enough such that it
could be extended to any cytoplasmic protein that prolongs the
open active conformation by binding to the active integrin.

Much as in the case of integrins, talins can be considered to
exist in two forms of different activity. Cytoplasmic talin is
inactivated by self-interactions and only the 'open' conformation
can bind to the membrane (PIP2), integrins and other proteins.
Therefore we can again simplify the model by only considering the
open active form, whose concentration can again be taken to be
constant, since any open talin bound to integrins can be expected
to be rapidly replenished from the pool of closed talins. Given
the low abundance of both open talin and integrin, complexes of
the two are taken to be absent in the absence of ligand. The gains
and losses of the state with $i$ integrins and $t$ talins bound
can again be translated into the linear Master equation

\begin{eqnarray}\label{Master_eq}
\dot{p}_{i,t} &=& k_{-} (i+1-t) p_{i+1,t}+k_{-}^* (t+1)p_{i+1,t+1}\\
&& +k_+(L-i+1) p_{i-1,t} \nonumber\\
&& +l_{-}(t+1)p_{i,t+1}+ l_{+}(i-t+1)p_{i,t-1} \nonumber \\
&& -(k_{-}(i-t)+k_{-}^*t+k_+(L-i)+l_{+}(i-t)+l_{-}t)p_{i,t}
\nonumber
\end{eqnarray}

with $p_{i,t} = 0$ if $t>i$; $t,i<0$; $t,i>L$. Here $p_{i,t}$ is
the probability that $i$ (open) integrins and $t$ (open) talins
are bound to the ligand with $L$ sites. $p_{i,t} = 0$ if $t>i$
reflects the fact that talins can only attach to the ligand
indirectly by binding to integrins. $k_{+},k_{-},l_{+},l_{-}$
refer to the on- and off-rates of the integrins and talins
respectively. The on-rates include the density of integrin and
talin which are taken to be constant (see above). $k_{-}^*$ refers
to the integrin off-rate when talin is bound ($k_{-}^* \leq k_-$).

\begin{figure} [!h]
\linespread{1.3}
\begin{center}
\includegraphics[width=7cm]{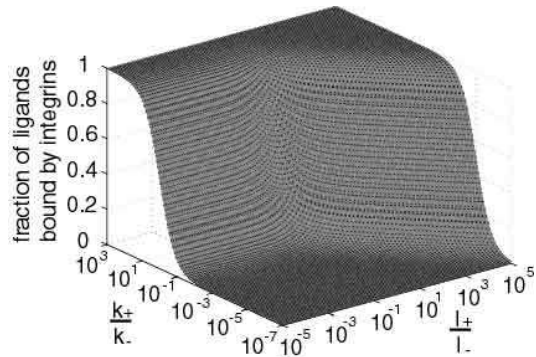}
\caption{\label{Fig_L_T} {\small \textbf{A stabilizing factor is not
sufficient for ligand-dependent integrin activation} The fraction of
ligand bound by integrin, $\frac{<I_b>}{L}$, is plotted against the
integrin-ligand association constant $K_a=\frac{k_+}{k_-}$ and the
integrin-talin association constant $K_a=\frac{l_+}{l_-}$ for $k_-^*
= 0$; L=10. For experimental estimates of these association
constants the activated fraction is small.}}
\end{center}
\end{figure}

For the steady state ($\dot{p}_{i,t}=0$) we obtain from
(\ref{Master_eq})

\begin{equation}\label{steadystate_talin}
p_{i,t} = \frac{\left(\begin{array}{c} L \\ t
\end{array}\right) \left(\begin{array}{c} L-t \\ i-t
\end{array}\right) a^{L-i}
b^{i-t}}{\sum_{i=0,t=0}^{i=L,t=L} \left(\begin{array}{c} L \\ t
\end{array}\right) \left(\begin{array}{c} L-t \\ i-t
\end{array}\right) a^{L-i}
b^{i-t}}
\end{equation}

with $a = \frac{k_{-}(k_{-}^*+l_{-})+l_{+} k_{-}^*}{l_{+} k_{+}}$,
$b = \frac{k_{-}^*+l_{-}}{l_{+}}$ for $i \geq t$ and $p_{i,t} =0$
for $i < t$. The expectation value for the number of integrins and
talins bound to a given ligand with L binding sites is given by
$<I_b> = \sum_{i=0}^{L} i \sum_{t=0}^{L} p_{i,t} $ and $<T_b> =
\sum_{t=0}^{L} t \sum_{i=0}^{L} p_{i,t}$ respectively. For strong
ligand binding in the absence of talin binding ($p_{L,0} \gg
p_{0,0}$) we require $b > a$ and we recover the condition $k_+ \gg
k_-$ found in the previous section. If talin binds the condition
becomes $\sum_{t=0}^{t=L} p_{L,t} \gg p_{0,0}$ and thus $a<b+1$.
This condition can be met if either $a<b$ or $a<1$. Assuming that
the integrin-ligand off-rate in the presence of talin is very
small ($k_-^* \ll 1$), $a<b$ still requires $k_{-} < k_{+}$ (high
integrin-ligand affinity) which we have found above to disagree
with experimental estimates. The condition $a<1$ can be met if
either $k_{-} \ll k_{+}$ or if $k_{-} > k_{+}$ and $l_{-} \ll
l_{+}$, that is integrin activation would be possible despite a
low integrin-ligand affinity if the integrin-talin affinity were
sufficient high (Fig. \ref{Fig_L_T}). However, such high
integrin-talin affinity would also lead to integrin activation in
the absence of ligand binding. We can therefore conclude that a
stabilizing factor alone is not sufficient for ligand-dependent
integrin activation.

\subsection*{A positive-feedback is required for ligand-dependent integrin activation}
The remaining two mechanisms in Fig. \ref{Fig_model} both involve
a positive feedback. Here, Figure \ref{Fig_model}C considers a
feed-back mechanism that only involves the integrin-ligand pair
itself while according to Figure \ref{Fig_model}D a larger network
would be necessary. Both mechanisms are analysed in the following.

\subsubsection*{A positive feedback that is based on the integrin-ligand pair} A positive feedback only
involving the integrin-ligand pair (Fig. \ref{Fig_model}C) could
either be enabled by integrin-integrin interactions in the open
form \cite{Li03} or by a ligand-induced conformational change that
leads to a higher affinity of binding. In both cases this positive
feedback ought to be triggered once a certain (small) number of
integrins is engaged in close proximity. Such interaction can thus
be captured in the model by replacing $k_-$ with $k_-^f$ in
(\ref{Master_eq_integrin}) and setting $k_-^f =
\frac{k_-^o}{(1+\frac{k_c i^n}{(i+K)^n})}$, where $k_-^o$ refers
to the integrin-ligand off-rate in the absence of a positive
feed-back. $k_c$ determines the strength with which the feedback
reduces the integrin-ligand off-rate, and $K$ and $n$ limit the
effect of integrin-integrin interactions to local interactions.
The model does not contain any spatial information and it is thus
assumed that integrins either bind to pre-clustered ligand, or
preferentially in the vicinity of other bound integrins. Note that
the assumption that ligand-integrin interactions are short-lived
in the absence of integrin-integrin interaction will lead to such
preferential binding in the vicinity of bound integrins. For the
steady state ($\dot{p}_{i}=0$) we then obtain from
(\ref{Master_eq_integrin})

\begin{equation}\label{steadystate_J}
p_{i} = \frac{\left(\begin{array}{c} L \\ i
\end{array}\right)\left(\frac{k_+}{ k_-}\right)^{i} \prod_{j=0}^i
\left(1+\frac{k_c j^n}{(j+K)^n}\right) }{\sum_{l=0}^L
\left(\begin{array}{c} L \\ l \end{array}\right) \left(\frac{k_+}{
k_-}\right)^{l} \prod_{j=0}^l {(1+\frac{k_c j^n}{(j+K)^n})}}.
\end{equation}

Given that $K$ and $n$ need to be chosen such that only local
integrin-integrin interactions reduce the ligand-integrin
off-rate, for sufficiently large $k_c$ (\ref{steadystate_J}) can
be approximated by

\begin{equation}\label{steadystate_J2}
p_{i} = \frac{\left(\begin{array}{c} L \\ i
\end{array}\right)\left(\frac{k_+ k_c}{
k_-}\right)^{i}}{\sum_{l=0}^L \left(\begin{array}{c} L \\ l
\end{array}\right)\left(\frac{k_+ k_c}{
k_-}\right)^{l}}
\end{equation}

(\ref{steadystate_J2}) and (\ref{steadystate}) are similar and
only differ in the factor $k_c$. The same argument thus applies
such that $k_c \gg \frac{ k_-}{k_+} \sim 10^2-10^3$ is required to
enable ligand-dependent integrin activation (Fig \ref{Fig_L}). In
the analysis of (\ref{steadystate}) $k_+$ was determined for the
high affinity integrin-ligand interaction and a further
ligand-induced affinity increase by 100-1000 fold is impossible.
Integrin-integrin interactions strong enough to reduce the high
affinity dissociation constant by a further factor of $10^2-10^3$
are also unlikely. Thus a feedback loop only involving integrins
is unlikely to drive ligand-dependent integrin activation.

\subsubsection*{A positive feedback loop based on a regulatory network can enable
ligand-dependent integrin activation} The last mechanism (Fig.
\ref{Fig_model}D) to be analysed is one that considers a
regulatory network that mediates the positive feedback. This
mechanism involves a stabilizing factor whose activity is
increased in response to ligand binding. Such positive feedback
has indeed been reported in the form that ligand-bound integrins
as well as talins trigger an increase in PIP2, which in turn
increases the recruitment of talin to the membrane, and thereby
the talin-integrin complex formation ($l_{+}$)
\cite{Garcia-Alvarez03,Martel01}. The talin density on the
membrane will thus be very small when integrins are inactive and
increase when ligands increase the integrin activity. This
positive feedback can be incorporated into the model as

\begin{equation}\label{lp}
l_{+} = l_c \frac{(t+1)^n}{(t+1)^n+K^n},
\end{equation}

such that the talin on-rate $l_{+}$ depends on the number of
talins in the integrin-ligand complex with $K$ being the Hill
constant, $n$ the Hill coefficient and $l_c$ some proportionality
factor. While many other formulations of such feedback are
possible, this saturation form captures the likely talin
dependence of the membrane PIP2 concentration, which will be low
below a certain talin threshold and eventually become saturated.

\begin{figure} [!h]
\linespread{1.3}
\begin{center}
\includegraphics[width=7cm]{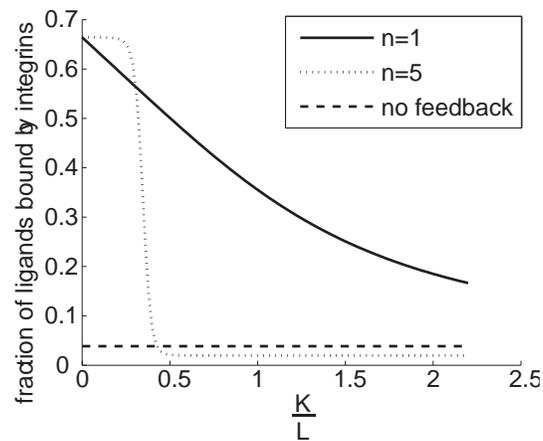}
\caption{\label{Fig_pos_feedback} {\small \textbf{A positive
feedback enables ligand-dependent integrin activation} The fraction
of ligand bound by integrin, $\frac{<I_b>}{L}$, is plotted in the
absence (- -) or presence of a talin feedback regulation against the
Hill constant $K$ (Eq. \ref{lp}) normalized by the number of binding
sites $L$ for different Hill coefficients [$n=1$ ($-$); $n=5$
($\cdots$)]. Parameters were set to $L=10$, $k_{+}=0.1$, $l_c=10^2$,
$k_{-}=5$, $k_{-}^* = 10^{-3}$, $l_{-}=1$. It can be seen that as
long as positive feedback is initiated upon binding of few integrins
(small $\frac{K}{L}$), a large active integrin fraction is
obtained.}}
\end{center}
\end{figure}

The steady state for $p_{i,t}$ is now more difficult to derive but
numerical studies show that such positive feedback indeed enables
ligand-induced integrin activation as long as $K/L \ll 1$ (Fig.
\ref{Fig_pos_feedback}). Note that, as before in the case of
integrin-integrin interactions, ligand-dependent integrin
activation requires the proportionality factor $l_c$ to be of
order $10^2-10^3$. In case of talin recruitment this is reasonable
since local PIP2 production may lead to a 100-1000 fold increase
in the local membrane talin density. A ligand-independent increase
in $l_{+}$, due to a signaling dependent increase in a factor that
stabilizes the open integrin conformation, will also lead to
integrin activation. This is likely to provide the mechanistic
basis for inside-out signaling.

We thus conclude that ligand-dependent integrin activation
requires $k_{-} \gg k_{+}$, $l_{-} \gg l_{+}$ in the resting and
$k_{-} \ll k_{+}$, $l_{-} \ll l_{+}$ in the active state. The
first condition is met by the binding of the stabilizing factor
which reduces $k_{-}$, and the second condition is enabled by a
positive feedback that leads to an increases in $l_{+}$ in the
presence of ligand.

\section*{Conclusion}
The analysis of possible mechanisms for integrin activation
suggests that a  positive feedback is required for
ligand-dependent integrin activation. Thus the ligand-integrin
affinity appears to be too low to stabilize the active integrin
conformation in the absence of a further stabilizing factor such
as talin. To ensure that ligand-independent integrin activation by
such a stabilizing factor alone is impossible a positive feedback
that upregulates the stabilizing factor upon ligand binding is
required.

This conclusion strongly depends on the order of magnitude of the
conversion factor $\eta$ between the 2- and 3-dimensional
dissociation constants and further careful measurements of this
parameter will be important. While ligand-dependent integrin
activation would be possible in the absence of further stabilizing
interactions if the theoretical rather than the experimental
estimates for $\eta$ were correct (that is if $\eta$ were 1000
times smaller), available experimental information is very much in
favour of the experimental estimate that has been used in this
study. In addition to the argument already given (see above)
further experimental observations corroborate the notion that the
physiological integrin-ligand affinity is tuned such that integrin
activation depends on supporting factors. Thus mutations in the
$\beta$-tail that disrupt the talin-integrin interaction markedly
reduce the fraction of active integrins \cite{Tadokoro03}. This
reduction can be overcome by binding of an antibody that
stabilizes the high affinity conformation \cite{Tadokoro03} which
is in agreement with the model prediction that a higher
ligand-integrin affinity can enable talin-independent integrin
activation. With integrin activation balanced on a knife-edge
small changes in the affinity, density and conformational bias of
integrins can have a large impact on the cell's adhesiveness and
motility - as is observed in experiments \cite{Palecek97}.

An advantage of a positive feedback regulation is the regulatory
potential. Signaling processes can affect the positive feedback,
providing a basis for inside-out signaling. It is likely that the
observed integrin clustering is a result of such a positive
feedback mechanism; a local increase in the density of a
stabilizing factor will facilitate the formation of further
integrin-ligand bonds in the vicinity of existing ones. Clustering
would then result from a process that enables long-lived
integrin-ligand interactions but is not of itself essential. The
concept that ligand binding and clustering are independent
processes is in agreement with the observation in
\textit{Drosophila} that integrins can still bind to the
extracellular matrix in the absence of talin but fail to cluster
\cite{Brown02}.

The feedback loop that is considered in the current model only
comprises a stabilizing factor (such as talin) and a way of
increasing the concentration of the stabilizing factor, such as by
PIP2 formation. Many more players are known to be involved and
understanding their relative contributions will be important.
Here, especially the role of PIPKI$\gamma$ will be of interest.
While PIPKI$\gamma$ produces PIP2 and thus increases talin
recruitment to the membrane, PIPKI$\gamma$ itself is recruited to
the membrane by talin \cite{Paolo02} and competes with the
integrin $\beta$ tail for talin binding \cite{dePereda05}.
Integrin activation leads to FAK activation and a FAK-enhanced
Src-mediated phosphorylation of PIPKI$\gamma$ further increases
its affinity for talin \cite{Ling02,Arias-Salgado03}.
PIPKI$\gamma$ therefore appears to play an important role both in
integrin activation and the turnover of focal contacts. A
combination of modeling and experiment, of the sort described
here, is expected to shed more light on this complex regulatory
network.

\paragraph{Acknowledgements}  We thank D.A. Calderwood and N.V. Valeyev for discussions.
D.I. is a Junior Research Fellow at St John's College, Oxford
University and is supported by an \mbox{EPSRC} scholarship. I.D.C
acknowledges financial support from the Wellcome Trust and the NIH
funded Cell Migration Consortium.


\newpage

\begin{figure} [!c]
\linespread{1.3}
\begin{center}
\includegraphics[width=10cm]{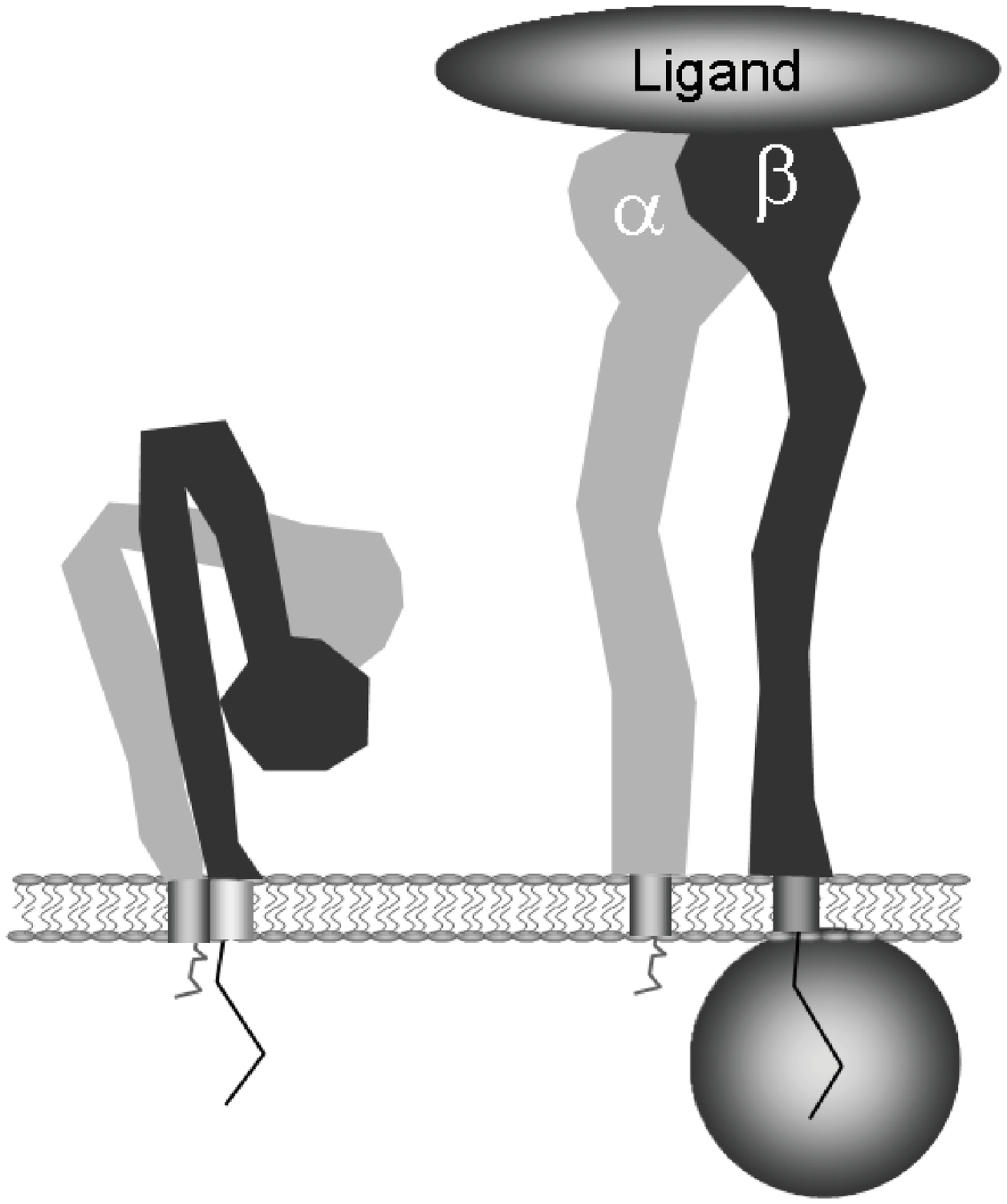}
\caption{\label{Fig_integrin} {\small \textbf{A simple model of
integrin activation.} Integrins exist in either a closed (LHS) or
open conformation (RHS); the open, high affinity form, can be
stabilized by binding of extracellular ligand or intracellular
proteins. The open conformation triggers downstream signaling and is
termed active.}}
\end{center}
\end{figure}

\begin{figure} [!c]
\linespread{1.3}
\begin{center}
\includegraphics[width=8cm]{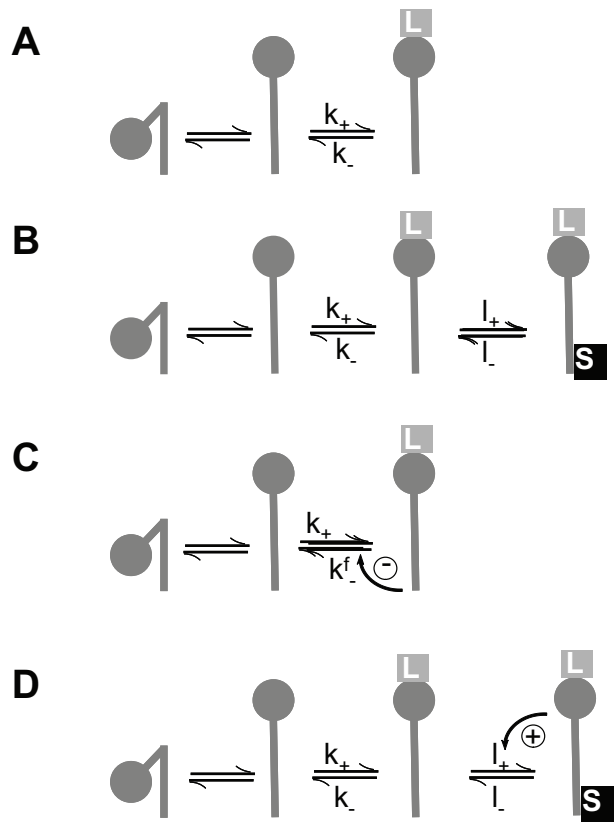}
\caption{\label{Fig_model} {\small \textbf{Four possible mechanisms
for integrin activation.} \textbf{(A)} Ligand-binding is sufficient
for integrin activation; other factors are only important for
down-stream signaling events. \textbf{(B)} A stabilizing factor
(such as talin) is necessary in addition to ligand binding.
\textbf{(C)} A ligand induced positive feedback (based for instance
on integrin self-interaction) induces integrin activation.
\textbf{(D)} A positive feedback involving further signaling
molecules is necessary for integrin activation. The symbols
represent simple variants of those in Figure \ref{Fig_integrin}. $L$
refers to ligand, $S$ to a stabilizing factor. The rate constants
correspond to those used in the models.}}
\end{center}
\end{figure}

\end{document}